\documentclass[fleqn,twoside]{article}
\usepackage{espcrc2}
\usepackage{graphicx}
\usepackage{amsmath}
\usepackage{mathtext}
\usepackage{amssymb}

\newcommand{\beq}{\begin{equation}}
\newcommand{\eeq}{\end{equation}}
\newcommand{\beqn}{\begin{equation*}}
\newcommand{\eeqn}{\end{equation*}}

\newcommand{\cA}{{\cal A}}

\newcommand{\GeV}{~\text{GeV}}

\title{
\vspace{-18mm}
\thispagestyle{empty}
\rightline{\normalsize ITEP-LAT/2002-12}
\vspace{-2mm}
\rightline{\normalsize KANAZAWA-02-25}
\vspace{-2mm}
\rightline{\normalsize {\bf 3} September, 2002}
Gluon propagators in maximal abelian gauge of $SU(2)$ lattice gauge theory
\thanks{Talk given by S.M.M. 
at Lattice 2002, Boston.
}}

\author{V.G.~Bornyakov \address[KU]{Institute for Theoretical Physics, Kanazawa
University, Kanazawa 920-1192, Japan}$^,$\address[ITEP]{Institute for Theoretical and Experimental
Physics, B.Cheremushkinskaya 25, Moscow, 117259, Russia}$^,$\address{Institute for High Energy Physics, Protvino, 142280, Russia},
S.M.~Morozov \addressmark[ITEP], M.I.~Polikarpov \addressmark[ITEP]}

\begin{document}

\begin{abstract}
We study propagators of diagonal and off-diagonal gluons in the momentum
space in maximal abelian gauge of $SU(2)$ lattice gauge theory. 
Remaining $U(1)$ degrees of freedom are fixed using Landau gauge. 
We find substantial 
difference between the propagator of the diagonal and the off-diagonal gluon
in the infrared region. The propagator of the off-diagonal gluon is
suppressed in comparison with that of the diagonal gluon at small momenta.
In the ultraviolet region both propagators behave as in nonabelian
Landau gauge.
\vspace{1pc}
\end{abstract}

\maketitle

\section{INTRODUCTION}

In lattice numerical studies gauge invariant quantities are usually computed.
On the other hand, gauge covariant quantities also provide important 
information.
The well known examples are quark and gluon propagators,
requiring complete gauge fixing,  monopoles and P-vortices, which study
needs only partial gauge fixing.
The first lattice calculations of the 
gluon propagator were performed in Landau gauge \cite{Mandula:rh}. 
Nowadays these results are significantly improved , also various gauges are
used (see e.g. \cite{will}).
Maximal abelian gauge (MAG), used to demonstrate the dual super conductor 
 confinement mechanism, is especially interesting.
Propagators in this gauge 
were not explored carefully enough so far. The first such study of propagators
in the coordinate space was performed in Ref.~\cite{Amemiya:zf}.
There was no study of propagators
in the momentum space. In this paper our aim is to close this gap.
We present our results of the high statistics calculation of propagators of
the diagonal and the off-diagonal gluon
in $SU(2)$ lattice gauge theory in MAG. Complete gauge fixing is 
achieved by using abelian Landau gauge to fix remaining abelian gauge 
degrees of freedom.

\section{GAUGE FIXING}

We use the standard parameterization of $SU(2)$ link matrices, 
$U_{11}=e^{i\theta}\cos\varphi,\ U_{12}=e^{i\chi}\sin\varphi$.
Then gauge fields are defined as follows:
\begin{align}
\label{eq:fields}
	A^1_\mu(x)&= \sin\varphi_\mu(x)\sin\chi_\mu(x),\\
	A^2_\mu(x)&= \sin\varphi_\mu(x)\cos\chi_\mu(x),\\
	A^3_\mu(x)&= \cos\varphi_\mu(x)\sin\theta_\mu(x).
\end{align}
We call $A_\mu^3(x)$ the diagonal gluon field, and $A_\mu^i(x),\ i = 1,2$ 
the off-diagonal gluon field.

\indent
The maximal abelian gauge condition in a differential form is
$$
(\partial_\mu \mp i A^3_\mu(x)) A^{\pm}_\mu(x)=0;\
A^{\pm}_\mu=\frac{1}{\sqrt{2}}(A^1_\mu \pm i A^2_\mu).
$$ 
Nonperturbative fixing of this gauge amounts to the minimization
of the functional
$$
F[A] = \int d^4 x ((A^1_\mu(x))^2 + (A^2_\mu(x))^2).
$$ 
In our simulations the Simulated Annealing algorithm \cite{Bali:1996dm} with 
20 randomly generated gauge copies is employed  to
minimize the effect of Gribov copies. 

\indent
After MAG, only $U(1)$ degrees of freedom remain unfixed.
We fix them using $U(1)$ Landau gauge. The differential lattice gauge condition
we are using is
\begin{equation}
\label{eq:contland}
\Delta_\mu \sin\theta_\mu(x) = 0.
\end{equation}
This condition implies that the diagonal field $A_\mu^3$ is not transversal
at a finite lattice spacing.
Another gauge condition $\Delta_\mu A^3_\mu=0$ is not considered here 
and will be discussed elsewhere \cite{inprep}.
Condition (\ref{eq:contland}) is equivalent to the maximization of the 
functional 
$$
\label{eq:landaulatfunc}
F[\theta] = \sum_{x,\mu} \cos\theta_\mu(x),
$$
using only $U(1)$ gauge transformations.
A local maximization algorithm with 30 random gauge copies is used to
accomplish this task. As the "stop criterion" for the algorithm we use 
\begin{equation}
\label{eq:stopcrit}
\frac{F[\theta]_{new} - F[\theta]_{old}}{F[\theta]_{new}} < 
\varepsilon = 10^{-8}.
\end{equation}

\section{PROPAGATORS}

We calculate the diagonal propagator
\begin{equation}
\label{eq:diagprop}
D_{\mu\nu}^{diag}(p)=D^{33}_{\mu\nu}(p)=\langle\cA^3_\mu(k)\cA^3_\nu(-k)\rangle, 
\end{equation}
and the off-diagonal propagator
\begin{multline}
\label{eq:offdiagprop}
D_{\mu\nu}^{offdiag}(p)=D^{11}_{\mu\nu}(p)=D^{22}_{\mu\nu}(p)=\\
\langle\cA^{1,2}_\mu(k)\cA^{1,2}_\nu(-k)\rangle,
\end{multline}
where Fourier transform $\cA_\mu^i(k)$ is defined as 
$$
\cA_\mu^i(k)=\frac{1}{\sqrt{L^4}}
\sum_x e^{-ik_\nu x_\nu-\frac{i}{2}k_\mu}A_\mu^i(x),
$$
$$
k_\mu=\frac{2\pi n_\mu}{aL_\mu}, n_\mu = 0,...,L_\mu-1.
$$
The physical lattice momenta $p$ are related to $k$ as follows:
$$
\label{eq:latmom}
p_\mu=\frac{2}{a}\sin{\frac{a k_\mu}{2}}. 
$$
Since both diagonal and off-diagonal fields are not transversal the 
general structure of diagonal and off-diagonal propagators is
\begin{equation}
\label{eq:gendmunu}
D_{\mu\nu}(p)=(\delta_{\mu\nu} - \frac{p_\mu p_\nu}{p^2})D^t(p^2) + 
\frac{p_\mu p_\nu}{p^2}D^l(p^2).
\end{equation}
Thus we have four structure functions $D^{t,l}_{diag,offdiag}$, 
which are really not independent. Note that in nonabelian Landau gauge,
$\partial_\mu A_\mu^a \tau^a=0$,
there would be only one formfactor:
$$
\langle\cA_\mu^a(p)\cA_\nu^b(-p)\rangle=\delta^{ab}(\delta_{\mu\nu} - 
\frac{p_\mu p_\nu}{p^2})D(p^2).
$$

\section{NUMERICAL RESULTS}
To calculate propagators 
the lattice with $L=24$ (138 configurations)
at $\beta = 2.40$ were simulated. The lattice spacing is $(1.66 \GeV)^{-1}$ at 
this $\beta$.

\begin{figure}[!thb]
\centering
\includegraphics[angle=-90,scale=0.3]{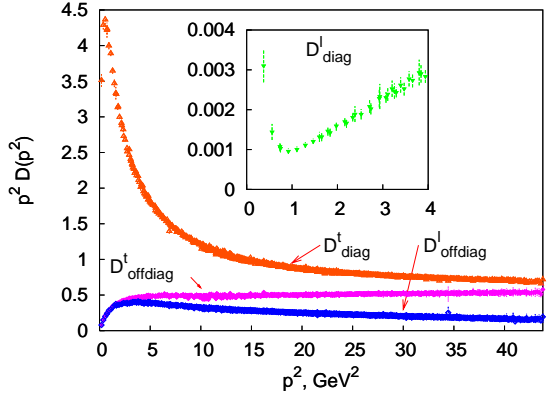}
\vspace{-8mm}
\caption{Formfactors of the diagonal and the off-diagonal gluon. The rescaled
inset shows $p^2D^l_{diag}$ with the same axes as in the main plot.}
\label{fig:both}
\vspace{-8mm}
\end{figure}

\indent
The behavior of transversal parts of propagators in the ultraviolet
region is the same as in nonabelian Landau gauge as our gauge corresponds
to that gauge in the limit of large momenta.
Both gluons are well described by the perturbative formula 
$D^t_{diag, offdiag}(p^2) = \frac{Z_{diag, offdiag}}{p^2}$, see
Fig.~\ref{fig:both}. 
The longitudinal part $p^2 D^l_{offdiag}(p^2)$ tends to zero as $p^2$ 
tends to infinity. And $p^2 D^l_{diag}(p^2)$ is small in comparison with other
structure functions (see inset in Fig. \ref{fig:both}). It increases linearly
with increasing $p^2$. 
Similar result was obtained in \cite{Chernodub:2001mg} for the 
longitudinal part of the
photon propagator in 3D compact QED when gauge condition was chosen allowing 
nonzero longitudinal part at the finite lattice spacing as in our case.  
Indications that the longitudinal part tends to zero in 
the continuum limit were found in  \cite{Chernodub:2001mg}.

\indent
The sharp increase
of $p^2D^l_{diag}(p^2)$ at low momenta seems to be related to imprecise
$U(1)$ Landau gauge fixing. We repeated computations on our smaller $16^4$
lattices
(300 gauge field configurations)
with reinforced condition~(\ref{eq:stopcrit}): $\varepsilon$ decreased
down to $10^{-10}$. We found that
$p^2D^l_{diag}(p^2) \rightarrow 0,\ p^2 \rightarrow 0$, \cite{inprep}.

In IR region one can see a very strong suppression of $D^t_{offdiag}(p^2)$ in
comparison with $D^t_{diag}(p^2)$ demonstrating the essence of the 
Abelian dominance 
\cite{Suzuki:1989gp}, Fig.\ref{fig:both}.
At the same time $D^l_{offdiag}(p^2)$ approaches $D^t_{offdiag}(p^2)$
at $p<1.5\GeV$. 
The off-diagonal propagator thus becomes:
\begin{equation}
\label{eq:offdiagfit}
D_{\mu\nu}^{offdiag}(p)=\frac{\delta_{\mu\nu}}{p^2+m^2_{off}(p^2)}.
\end{equation}
Fit to this form at $p<1.5\GeV$ with $\chi^2/N_{dof} = 1.76$ 
gives $m_{off}=1.06(1)\GeV$.

\indent
In the range $p<1.5\GeV$ the best fit ($\chi^2/N_{dof} = 0.51$)
to $D^t_{diag}(p^2)$ is given by the formula
\begin{equation}
\label{eq:diagfit}
D^t_{diag}(p^2) = \frac{Z\ m_{diag}^{2 \alpha}}
{(p^2+m^2_{diag})^{1+\alpha}}.
\end{equation}
We obtain $\alpha = 0.80(1),\ m_{diag}=0.63(1)\GeV$.
This shows that behavior of $D^t_{diag}(p^2)$ in the infrared region is qualitatively 
similar to that of the propagator in nonabelian Landau gauge~\cite{will}.
\begin{figure}[!thb]
\vspace{-8mm}
\centering
\includegraphics[angle=-90,scale=0.3]{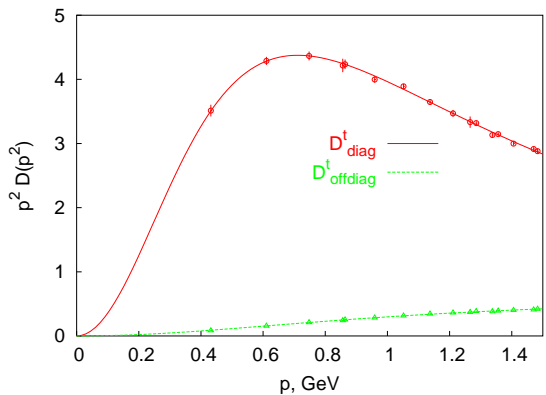}
\vspace{-8mm}
\caption{Lattice data fitted by Eq.~\ref{eq:diagfit} for the $D^t_{diag}(p^2)$ and by
 Eq.~\ref{eq:offdiagfit} for $D^t_{offdiag}(p^2)$.}
\label{fig:inverse}
\vspace{-8mm}
\end{figure}

\section{CONCLUSIONS AND OUTLOOK}
Our results clearly show that
off-diagonal gluons are suppressed at low momenta thus
providing the explanation of the Abelian dominance established in
numerical studies of MAG.
Effects of the finite volume and incomplete gauge fixing should be further
investigated.

\section*{ACKNOWLEDGMENTS}

The authors are grateful to F.V.~Gubarev, M.N.~Chernodub, A.~Schiller,
G.~ Schierholz and
V.I.~Zakharov for useful discussions.  M.~I.~P is partially supported by grants
RFBR 02-02-17308, RFBR 01-02-117456, RFBR 00-15-96-786, INTAS-00-00111, and
CRDF award RPI-2364-MO-02.
S.~M. is partially supported by grants RFBR 02-02-17308 and CRDF MO-011-0.
V.G.B. is supported by JSPS Fellowship.

\end{document}